\documentstyle[12pt,psfig]{article}
\newcommand {\beq}{\begin{equation}}
\newcommand {\eeq}{\end{equation}}
\newcommand {\beqa}{\begin{eqnarray}}
\newcommand {\eeqa}{\end{eqnarray}}
\newcommand {\beqan}{\begin{eqnarray*}}
\newcommand {\eeqan}{\end{eqnarray*}}
\newcommand {\n}{\nonumber \\}

\newcommand {\Romannumeral}[1]{\uppercase\expandafter{\romannumeral#1}}

\newcommand {\barU}{\bar U}

\newcommand {\mw}{\mbox{\scriptsize (Maj)}}
\newcommand {\MW}{\mbox{M}}
\newcommand {\wb}{\mbox{\scriptsize WB}}
\newcommand {\adj}{\mbox{\scriptsize (A)}}
\newcommand {\ham}{\mbox{H}_\pm}
\newcommand {\mbham}{{\cal H}_\pm}
\newcommand {\wilson}{\mbox{\bf B}}
\newcommand {\chiral}{\mbox{\bf C}}
\newcommand {\tr}{\mbox{tr}}
\newcommand {\diag}{\mbox{diag}}
\newcommand {\ee}{\mbox{e}}

\newcommand {\dd}{\mbox{d}}

\begin{document}
\setlength{\oddsidemargin}{0cm}
\setlength{\baselineskip}{7mm}

\begin{titlepage}
\renewcommand{\thefootnote}{\fnsymbol{footnote}}
    \begin{normalsize}
     \begin{flushright}
                 DPNU-97-32\\
                 November 1997
     \end{flushright}
    \end{normalsize}
    \begin{Large}
       \vspace{1cm}
       \begin{center}
         {Unitary IIB Matrix Model}\\
{and the Dynamical Generation of the Space Time} \\
       \end{center}
    \end{Large}

  \vspace{10mm}

\begin{center}
           Naofumi K{\sc itsunezaki}\footnote
           {E-mail address : kitsune@eken.phys.nagoya-u.ac.jp
} {\sc and}
           Jun N{\sc ishimura}\footnote
           {E-mail address : nisimura@eken.phys.nagoya-u.ac.jp}\\
      \vspace{1cm}
{\it Department of Physics, Nagoya University,} \\
              {\it Chikusa-ku, Nagoya 464-01, Japan}\\
\vspace{15mm}

\end{center}
\hspace{5cm}

\begin{abstract}
\noindent
We propose  a unitary matrix model as 
a regularization of the IIB matrix model of
Ishibashi-Kawai-Kitazawa-Tsuchiya (IKKT).
The fermionic part is incorporated using the overlap formalism
in order to avoid unwanted ``doublers'' while preserving the
global gauge invariance.
This regularization, unlike the one adopted by IKKT,
has manifest U(1)$^{10}$ symmetry, which 
corresponds to the ten-dimensional translational invariance of the
space time.
We calculate one-loop effective action around some typical
BPS-saturated configurations in the weak coupling limit.
We also discuss a possible scenario for the 
dynamical generation of the four-dimensional space time 
through spontaneous breakdown of the U(1)$^{10}$ symmetry in the 
double scaling limit.
\end{abstract}

\end{titlepage}

\vfil\eject

\setcounter{footnote}{0}

\section{Introduction}
\setcounter{equation}{0}

String theory has been studied as a natural candidate of the unified theory
including quantum gravity.
As the theory of everything, it should explain all the details of the 
standard model as low energy physics, such as
the structure of the gauge group, the three generations of the matter,
and even the space-time dimension.
Perturbative study of string theory in the eighties revealed, however, that 
there are infinitely many perturbatively stable vacuua, 
and that we cannot make any physical prediction as to our present world 
unless we understand the nonperturbative effects.
It is natural to expect that just as the confinement in QCD was
understood only after lattice gauge theory appeared as a constructive 
definition of gauge theories \cite{Wilson}, so must the true vacuum of 
superstring theory be understood once a constructive definition of string 
theories could be obtained.
Recently, such a constructive definition of superstring theory
using large $N$ matrix model \cite{BFSS,IKKT} has been proposed.
Solitonic objects known as D-branes \cite{Polch,Witten}, which was focused
in the context of string duality, plays an important role here.
The basic idea is to quantize the lowest dimensional D-brane 
nonperturbatively, instead of string itself.
It has been shown \cite{FKKT} that 
$1/N$ expansion of the model proposed 
by Ishibashi-Kawai-Kitazawa-Tsuchiya (IKKT) \cite{IKKT}
gives the perturbation theory of type IIB superstring
by reproducing the light-cone string field theory through 
Schwinger-Dyson equation.
Above all, the way in which one should take the double scaling limit
has been explicitly identified at least for sufficiently small string 
coupling constant.
We may say that we are now at the stage to extract nonperturbative 
physics of superstring through this model.

As one of the most fundamental issues,
let us consider how we can get the space-time dimension.
Since the eigenvalues of the bosonic hermitian matrices of the IKKT
model are interpreted
as space-time coordinates of the D-objects, 
one possible scenario for the dynamical
generation of the space time should be that in an appropriate double scaling
limit the distribution of the eigenvalues degenerates to a 
four-dimensional hypersurface.
Note here that the model before regularization possesses 
a symmetry under a transformation 
which shifts all the hermitian matrices by the identity matrix times
a constant,
which corresponds to the 10D translational invariance of the
eigenvalue space.
The degeneracy of the eigenvalue distribution, therefore, 
implies the spontaneous breakdown of the translational invariance.
The translational invariance is also essential in reproducing the string
perturbation theory \cite{FKKT}.
A regularization, however, adopted by IKKT \cite{IKKT} 
in order to make the integration
over the bosonic hermitian matrices well-defined by requiring that the 
magnitude of the eigenvalues should be less than $\pi/a$, 
clearly violates the 10D translational invariance at the boundary of 
the eigenvalue space.
A hope \cite{FKKT} might be that this does not cause any serious problem, 
since we take the cutoff $a$ to zero in the end anyway,
but it is certainly a flaw of this model.

We therefore consider in this paper,
a regularization which preserves manifest 10D translational invariance.
A natural candidate is to replace the bosonic Hermitian matrices by
unitary matrices
\footnote{A unitary matrix model for M(atrix) Theory 
has been considered in Ref. \cite{Poly} from a completely different 
motivation.}.
It is now the phases of the eigenvalues that are 
interpreted as the space-time coordinates.
Thus the space time is naturally compactified to a ten-dimensional torus.
The unitary matrix model can be considered as being reduced from 
a 10D lattice gauge theory.
An obvious drawback of this kind of model, therefore,
is that the continuous rotational invariance 
is broken down to the discrete one.
Moreover a problem related to lattice fermion arises here.
Namely when we consider the fermionic part naively, $2^{10}=1024$ doublers
will come out. Since the fermionic matrices are Majorana-Weyl spinor
as a representation of the Lorentz group, we cannot easily decouple
the unwanted doublers as is the case with ordinary lattice chiral
gauge theories.

The problem of regularizing chiral gauge theories on the lattice
has a long history and there are a lot of proposals made so far.
The most promising one at present is the overlap formalism \cite{NN},
which passed a number of tests \cite{Rajamani}.
The idea to apply this formalism to ten-dimensional ${\cal N}=1$
(anomalous and non-renormalizable) super Yang-Mills theory,
as a regularization of four-dimensional ${\cal N}=4$ super Yang-Mills
theory via dimensional reduction, is mentioned in Ref. \cite{MajoranaWeyl}.
What we should consider here is 
essentially the large $N$ reduced version of it.
We note that although the overlap formalism as a regularization of 
ordinary lattice chiral gauge theories has a subtle problem with
the local gauge invariance not being preserved on the lattice,
its application to the present case is completely safe regarding this,
since we do not have the local gauge invariance to take care of.
The global gauge invariance, on the other hand, 
which is indeed one of the important symmetry of the model,
is manifestly preserved within the formalism.
The unitary matrix model thus defined has the U(1)$^{10}$ symmetry,
which corresponds to the 10D translational invariacne of the 
space time.

Another important symmetry of the IKKT model is the ${\cal N}$=2 
supersymmetry regarding the eigenvalues of the bosonic Hermitian matrices 
as the space-time coordinates \cite{IKKT}.
Note that the 10D translational invariance mentioned above
gives a subgroup of the supersymmetry.
This ${\cal N}$=2 supersymmetry comes from (1) the supersymmetry 
of 10D super Yang-Mills theory, combined with (2) 
the symmetry under constant shifts
of the fermionic matrices.
When we consider the unitary matrix model with the overlap formalism,
we have (2) but not (1) unfortunately, and therefore,
the ${\cal N}$=2 supersymmetry is not manifest.
In the case of ordinary super Yang-Mills theories, 
supersymmetry is broken when we put the theory on the lattice,
but is expected to be restored in the continuum limit 
with appropriate fine-tuning, as has been advocated in Ref.~\cite{CV}.
Fine-tuning can be avoided by the use of the overlap formalism
due to the exact chiral symmetry 
for 4D ${\cal N}=1$ super Yang-Mills theory \cite{NN}
and due to the exact parity invariance 
for 3D ${\cal N}=1$ super Yang-Mills theory \cite{superYM}.
This is based on the universality argument.
Although the validity of the universality argument
in matrix models is not clear at present,
we naively expect the ${\cal N}$=2 supersymmetry to be restored in the 
double scaling limit without particular fine-tuning.

To summarize our strategy, we respect the U(1)$^{10}$ symmetry 
or the 10D translational invariance
as the most important symmetry of the model, while
we give up the continuous 10D Lorentz invariance and the ${\cal N}$=2 
supersymmetry
as manifest symmetries and naively expect them to be restored in the 
double scaling limit.
This we consider to be analogous to the case with lattice gauge theory 
\cite{Wilson},
where gauge symmetry is respected and the continuous Poincare
invariance is broken by the lattice regularization and 
expected to be restored only in the continuum limit.

We study our model through one-loop perturbative expansion 
in the weak coupling limit.
We calculate one-loop effective action around classical vacua.
In this case, 
the integration over the bosonic degrees of freedom 
is already done in Ref. \cite{BHN}
in the context of Eguchi-Kawai model \cite{EK},
and it is known to give rise to the logarithmic attractive potential
between two eigenvalue points \cite{BHN}.
This attractive potential is exactly cancelled
by the contribution from the fermionic degrees of freedom when
the eigenvalues are sufficiently close to each other
in accordance with the IKKT model \cite{IKKT}.
On the other hand, when the eigenvalues are farther apart, a weak attractive
potential arises, unlike the case with IKKT model, where the potential is
completely flat.
Thus the U(1)$^{10}$ symmetry of our model
is spontaneously broken in the weak coupling limit, 
though much more mildly than 
in purely bosonic case, where the logarithmic attractive potential 
exists \cite{BHN}.
We also calculate the effective action around other BPS-saturated
configurations which represent one D-string and two parallel D-strings.

Since the U(1)$^{10}$ symmetry is expected to be restored
in the strong coupling phase, there must be a phase transition.
We consider that this phase transition provides a natural place 
to take the double scaling limit.
Since our model preserves manifest U(1)$^{10}$ symmetry,
we can discuss the dynamical generation of the space time
as the spontaneous breakdown of the U(1)$^{10}$ symmetry.
Roughly speaking, if the U(1)$^{10}$ symmetry is broken down to
U(1)$^{D}$ for sufficiently small string coupling constant, 
our model is equivalent to a $D$-dimensional gauge theory
due to the argument of Eguchi-Kawai \cite{EK}.
However, according to Ref. \cite{FKKT}, we have to take the double
scaling limit at the weak coupling region.
Since gauge theories in more than four dimensions
should have an ultraviolet fixed point at the strong coupling regime, if any,
the U(1)$^{10}$ symmetry of our model is expected to be broken down 
to U(1)$^{4}$ at least for sufficiently small string coupling constant.
This gives a qualitative understanding of the dynamical origin of
the space-time dimension 4.


This paper is organized as follows.
In Section 2, we define our model and discuss the symmetries it possesses.
In Section 3, 
we calculate the one-loop effective action around some typical
BPS-saturated configurations including the one corresponding to 
the classical vacua, which shows that the U(1)$^{10}$ symmetry 
is spontaneously broken in the weak coupling limit. 
In Section 4, we consider the double scaling limit of the present
model.
We argue that the phase transition accompanied with 
the spontaneous breakdown of the
U(1)$^{10}$ symmetry provides a natural place to take the
double scaling limit
and that the dynamical generation of the four-dimensional space time 
could be naturally expected to occur at the critical region.
Section 5 is devoted to summary and future prospects.

\vspace{1cm}

\section{The Unitary IIB Matrix Model and Its Symmetries}
\setcounter{equation}{0}

\subsection{Review of the IKKT model}

In this paper, we work in Euclidean space time.
The IKKT model \cite{IKKT} is defined through the action
\beq
S= -\frac{1}{g^2} \left( \sum_{\mu \nu}
\frac{1}{4}\tr [A_\mu,A_\nu]^2 
+ \frac{1}{2}\sum_{\mu} \tr \overline{\psi} \Gamma_\mu [A_\mu, \psi] \right),
\label{eq:IKKTaction}
\eeq
which can be formally obtained 
by the zero-volume limit of 10D supersymmetric U($N$) Yang-Mills theory.
$A_\mu$ and $\psi$ are $N\times N$ hermitian matrices,
which transform as a vector and a Majorana-Weyl spinor respectively
under 10D Lorentz group.
$\Gamma_\mu$ are 10D gamma matrices which satisfies
\beq
\{ \Gamma_\mu, \Gamma_\nu \} = 2 \delta_{\mu\nu}.
\eeq

Let us briefly review the symmetries of this model.
The following three come from
the 10D Lorentz invariance, ${\cal N}$=1 supersymmetry, 
and the global gauge
invariance of the 10D supersymmetric Yang-Mills theory, respectively.

\noindent \underline{(i) 10D rotational invariance}

\noindent \underline{(ii) zero-volume version of the supersymmetry}
\beqa
\delta \psi &=& \frac{i}{2}  \sum_{\mu \nu}
[A_\mu,A_\nu] \Gamma_{\mu\nu} \epsilon, \n
\delta A_\mu &=& i \overline{\epsilon} \Gamma_\mu \psi,
\eeqa
where $\Gamma_{\mu\nu} = \frac{1}{2} (\Gamma_\mu \Gamma_\nu
- \Gamma_\nu \Gamma_\mu )$.

\noindent \underline{(iii) global gauge invariance}
\beqa
\delta \psi  &=& i [\psi,\alpha], \n
\delta A_\mu &=& i [A_\mu,\alpha],
\eeqa

Note that although 10D supersymmetric Yang-Mills theory has gauge
anomaly, this is not of much concern to us since the model after
the zero-volume limit do not have {\em local} gauge invariance anyway.
Note also that the IKKT model can actually be defined without any
reference
to 10D supersymmetric Yang-Mills theory, which cannot be considered
as a consistent quantum field theory due to the gauge anomaly.

In addition to these symmetries, the following symmetries arise
due to the zero-volume limit.

\noindent \underline{(iv) constant shift of the bosonic hermitian
matrices}
\beqa
\delta \psi &=& 0 \n
\delta A_\mu &=& \alpha_\mu
\eeqa

\noindent \underline{(v) constant shift of the fermionic hermitian matrices}
\beqa
\delta \psi = \xi \n
\delta A_\mu = 0
\eeqa

\noindent As is shown in Ref. \cite{IKKT}, 
the symmetries (ii) and (v),
with the aid of the symmetry (iii) and the equations of motion
give rise to the ten-dimensional 
${\cal N}=2$ supersymmetry, regarding the eigenvalues 
of the bosonic hermitian matrices as the space-time coordinates.
The symmetry (iv), which corresponds to the 10D translational invariance,
gives a subgroup of this supersymmetry.

The integration over the bosonic hermitian matrices has to be
regularized in order to be well-defined.
In Ref. \cite{IKKT}, the regularization was given by restricting the
integration region to
\beq
|\mbox{eigenvalues of the }A_\mu|\le \frac{\pi}{a},
\eeq
where $a$ is the cutoff.
Due to this regularization, the symmetries (i),(ii) and (iv) is violated
at the boundary of the eigenvalue space.
Therefore, the ten-dimensional ${\cal N}=2$
super Poincare invariance is not preserved in the strict sense.
Although one might hope that this does not cause any serious problem
since the cutoff is taken to zero with proper renormalization of the
coupling constant in the end, it is certainly a flaw of this model.

\subsection{Definition of the unitary IIB matrix model}

We consider a regularization that has manifest symmetries
which correspond to (iii), (iv), (v),
and the discrete subgroup of the rotational invariance (i).
Above all, as compared with the IKKT model, our model possesses
manifest U(1)$^{10}$ symmetry, which corresponds
to the ten-dimensional translational invariance
of the space time.
We replace the bosonic hermitian matrices of the IKKT model by 
unitary matrices.
The bosonic part of the action can be written as
\beq
S_b = - N \beta \sum_{\mu\nu} \tr
U_\mu U_\nu U_\mu^\dagger U_\nu^\dagger.
\label{eqn:S_b}
\eeq
The fermionic part could be naively written as
\beq
S_f = \sum_{\mu} \frac{1}{4} \left( \tr
\overline{\psi} \Gamma_\mu U_\mu \psi U_\mu^\dagger
- \tr  \overline{\psi} \Gamma_\mu U_\mu^\dagger \psi U_\mu \right).
\eeq
The total action $S_b+S_f$ can be formally obtained by reducing the 
10D lattice gauge theory with naive Majorana-Weyl fermion to
a unit cell.
The naive fermion action 
in ordinary lattice gauge theories in $D$ dimensions 
gives rise to $2^D$ doublers.
The doublers can be seen as
duplicated poles in the free fermion propagator in the momentum space.
In the present model, 
we can see them by considering the perturbative expansion around
the classical vacua
\beq
U_\mu = \diag(\ee^{i\theta_{\mu 1}},\ee^{i\theta_{\mu 2}},\cdots,
\ee^{i\theta_{\mu N}}).
\eeq
The fermion propagator is given by 
\beq
\langle \psi_{pq} \psi_{rs} \rangle
= \left( \sum_{\mu}\Gamma_\mu \sin (\theta_{\mu p} - \theta_{\mu
q}) \right)^{-1} \delta_{ps} \delta_{qr}.
\eeq
The $2^{10}$ poles are given by $\theta_{\mu p } - \theta_{\mu q} = 0$ or 
$\pi$ for each $\mu$, 
while the only pole that appears in the IKKT model is given by
$\theta_{\mu p } - \theta_{\mu q} = 0$.
Thus, unless the U(1)$^{10}$ is completely broken spontaneously
and $\theta_{\mu p } - \theta_{\mu q} \approx 0$ dominates,
the naive fermion action gives 1024 doublers, half of which being left 
handed and the rest being right handed.
This is unacceptable since it violates the balance between the bosonic
degrees of freedom and the fermionic degrees of freedom.
Since the fermions are chiral, it is not easy to decouple the
doublers, as is the case with ordinary lattice chiral gauge theories.

We overcome this problem in the present case, applying the overlap
formalism,
which has been developed to deal with lattice chiral gauge theories.
We note that although the overlap formalism, as a regularization 
of lattice chiral gauge theory, has a subtle problem related to the 
local gauge invariance on the lattice, in the present application,
we are completely free from this subtlety, since there is no local 
gauge invariance we have to respect.
On the other hand, the overlap formalism preserves
the 10D discrete rotational invariance and the global gauge invariance.
Also we can identify a zero mode that corresponds to
the fermion shift invariance.

We introduce the following notation for the adjoint representation
of the U($N$) group.
\beqa
\psi_{pq} &=& \sum_{a} (T^a)_{pq} \psi_a^{\adj} \n
\overline{\psi}_{pq} &=& \sum_{a} (T^a)_{pq} \overline{\psi}_a^{\adj},
\eeqa
where $T^a$ are the generators 
normalized as $\tr(T^a T^b)=\delta_{ab}$.
Then the fermion action can be written as
\beqa
S_f &=& \sum_{\mu} \frac{1}{4} \left( 
\overline{\psi}^{\adj} \Gamma_\mu U_\mu^{\adj} \psi^{\adj}
-  \overline{\psi}^{\adj} \Gamma_\mu U_\mu^{\adj\dagger} \psi^{\adj}
\right) \n
&=& 
 \sum_{\mu} \frac{1}{2} 
\overline{\psi}^{\adj} \Gamma_\mu D_\mu \psi^{\adj} ,
\label{eq:faction}
\eeqa
where we have introduced the adjoint link variable through
\beq
(U_\mu^{\adj})_{ab} = \tr (T^a U_\mu T^b U_\mu^\dagger)
\label{eq:adjointlink}
\eeq
and the reduced version of the covariant derivative through
\beq
D_\mu = \frac{1}{2} (U_\mu^{\adj}-U_\mu^{\adj\dagger}).
\eeq
One can see that $U_\mu^{\adj}$ is real, {\it i.e.},
$(U_\mu^{\adj})^*=U_\mu^{\adj}$, which reflects the fact that
the adjoint representation is a real representation.

We take a particular representation for the 10D 
gamma matrices in the following way.
We first define 8D gamma matrices $\gamma_i$ ($i= 1,2,\cdots, 8$)
which satisfies $\{ \gamma_i, \gamma_j\} = \delta_{ij}$.
We take $\gamma_9= \gamma_1 \gamma_2 \cdots \gamma_8$.
Then the 10D gamma matrices is defined as
\beqa
\Gamma_i &=& \sigma_1 \otimes \gamma_i ~~~~~~ (i= 1,2,\cdots, 9) \n
\Gamma_{10} &=& \sigma_2 \otimes {\bf 1}   .
\label{eq:10Dgamma}
\eeqa 
Note that the chirality operator 
\beqa
\Gamma_{11} &=& - i \Gamma_1 \Gamma_2 \cdots \Gamma_{10} \n
&=& \sigma_3 \otimes {\bf 1} 
\eeqa
is diagonal, which implies that 
the above construction gives a Weyl representation of the gamma
matrices.
Then the action for the fermionic matrices can be written as
\beqa
S_f &=& 
 \sum_{\mu} \frac{1}{2} 
\overline{\psi}^{\adj} \Gamma_\mu D_\mu \psi^{\adj} \n
&=& 
 \sum_{\mu} \frac{1}{2} 
\psi^{\adj\dagger} i \Gamma_{10}  \Gamma_\mu D_\mu \psi^{\adj} \n
&=&
 \frac{1}{2} 
\psi^{\adj\dagger} ( i D_{10} + 
\sigma_3 \otimes \sum_i \gamma_i D_i) \psi^{\adj} \n
&=&
 \frac{1}{2} 
\left(
\begin{array}{cc}
\psi_L^{\adj\dagger} & \psi_R^{\adj\dagger}
\end{array}
\right)
\left(
\begin{array}{cc}
\sum_i \gamma_i D_i + i D_{10}  & 0 \\
0 &   -  \sum_i \gamma_i D_i + i D_{10} 
\end{array}
\right)
\left(
\begin{array}{c}
\psi_L^{\adj}  \\
\psi_R^{\adj}  
\end{array}
\right)   \n
&=&
 \frac{1}{2} 
\left(
\begin{array}{cc}
\psi_L^{\adj\dagger} & \psi_R^{\adj\dagger}
\end{array}
\right)
\left(
\begin{array}{cc}
\chiral & 0 \\
0 &   \chiral^\dagger
\end{array}
\right)
\left(
\begin{array}{c}
\psi_L^{\adj}  \\
\psi_R^{\adj}  
\end{array}
\right),
\eeqa
where we define the chiral operator as
\beq
\chiral= \sum_\mu \gamma_\mu D_\mu 
= \sum_\mu \gamma_\mu \frac{1}{2} (U_\mu^{\adj}-U_\mu^{\adj\dagger}),
\eeq
and define $\gamma_{10}$ as
\beq
\gamma_{10} = i {\bf 1}.
\eeq
Note that $\gamma_i$ ($i=1,2,\cdots 9$) are hermitian, while
$\gamma_{10}$ is anti-hermitian.

Let us first consider the fermion determinant for a 10D Weyl fermion
in the present case through the overlap formalism \cite{NN}.
We consider the many-body Hamiltonians defined as
\beq
\mbham = 
\left(
\begin{array}{cc}
\alpha^{\dagger} & \beta^\dagger
\end{array}
\right)
\left(
\begin{array}{cc}
\wilson \pm m & \chiral  \\
\chiral ^\dagger & -(\wilson \pm m)
\end{array}
\right)
\left(
\begin{array}{c}
\alpha  \\
\beta
\end{array}
\right).
\label{eq:mbham}
\eeq
where $\{ \alpha^\dagger, \alpha \}$ and 
$\{ \beta^\dagger, \beta \}$
are sets of creation and annihilation operators
which obey the canonical anti-commutation relations.
The fermionic operators carry spinor indices as well as those for
the adjoint representation of the U($N$) group,
which we have suppressed.
$\wilson$ is defined through
\beq
\wilson= \sum_\mu 
\left\{ 1 - \frac{1}{2}(U_\mu^{\adj} + U_\mu^{\adj\dagger}) \right\},
\eeq
which plays the role of the Wilson term and eliminates the 
doublers \cite{NN}.
We will see this explicitly in the next section.
$m$ is a constant which should be kept fixed within $0 < m < 1$
when one takes the double scaling limit.

We denote the ground states of the many-body Hamiltonians $\mbham$
as $|\pm\rangle_U$.
We fix the $U$ dependence of the phase of the states by imposing
that $~_1\langle\pm|\pm\rangle_U$ should be real positive.
This is called the Wigner-Brillouin phase choice in Ref. \cite{NN}.
We denote the ground states thus defined as $|\pm \rangle _U ^{\wb}$.
Then the fermion determinant for a single 
left-handed Weyl fermion is defined as 
$~^{\wb}_{~~U}  \langle - | + \rangle _U ^{\wb} $.

We can see that the above formula 
can be decomposed into two Majorana-Weyl
fermions following the steps taken in Ref. \cite{MajoranaWeyl}.
Note first that 
there exists a unitary matrix $\Omega$ which has the following
property.
\beqa
\Omega^\dagger \gamma_i \Omega &=& (\gamma_i)^t~~~~~(i=1,2,\cdots,10) 
\label{eq:conj}
\\
\Omega^t &=& \Omega.
\label{eq:trans}
\eeqa
For example, one can take the unitary matrix $\Omega$ 
to be identity, by choosing
all the $\gamma_i$ ($i=1,2,\cdots, 9$) to be real,
which corresponds to the case in which
$\Gamma_{10}$ and $i\Gamma_i $ ($i=1,2,\cdots,9$)
defined through eq.(\ref{eq:10Dgamma})
give a Majorana-Weyl representation of 
the 10D gamma matrices in Minkowski space.
Using eqs. (\ref{eq:conj}) and (\ref{eq:trans}), one can show that
\beq
(\chiral \Omega)^t = - \chiral \Omega.
\eeq 
We make the following Bogoliubov transformation.
\beqa
\alpha &=& \frac{\xi+i\eta}{\sqrt{2}} \n
\beta &=& \Omega \frac{\xi^{\dagger t} + i \eta^{\dagger t} }{\sqrt{2}}.
\eeqa
Plugging this into eq.(\ref{eq:mbham}),
we obtain 
\beqa
\mbham &=& 
\frac{1}{2}
\left(
\begin{array}{cc}
\xi^{\dagger} & \xi^t
\end{array}
\right)
\left(
\begin{array}{cc}
\wilson \pm m & \chiral \Omega \\
(\chiral \Omega) ^\dagger & -(\wilson \pm m)
\end{array}
\right)
\left(
\begin{array}{c}
\xi  \\
\xi^{\dagger t}  
\end{array}
\right)  \n
&~&+
\frac{1}{2}
\left(
\begin{array}{cc}
\eta^{\dagger} & \eta^t
\end{array}
\right)
\left(
\begin{array}{cc}
\wilson \pm m & \chiral \Omega \\
(\chiral \Omega) ^\dagger & -(\wilson \pm m)
\end{array}
\right)
\left(
\begin{array}{c}
\eta  \\
\eta^{\dagger t}  
\end{array}
\right).
\eeqa
Thus, $\xi$ and $\eta$ decouple and 
each term of the Hamiltonians corresponds to a Majorana-Weyl fermion.

The fermion determinant for a Majorana-Weyl fermion can,
therefore, be defined as the overlap 
of the ground states of the two many-body Hamiltonians:
\beq
\mbham^{\mw} = 
\frac{1}{2}
\left(
\begin{array}{cc}
\xi^{\dagger} & \xi^t
\end{array}
\right)
\left(
\begin{array}{cc}
\wilson \pm m & \chiral \Omega \\
(\chiral \Omega)^\dagger & -(\wilson \pm m)
\end{array}
\right)
\left(
\begin{array}{c}
\xi  \\
\xi^{\dagger t}  
\end{array}
\right).
\label{eq:mbhamMW}
\eeq
Let us denote the ground states with the Wigner-Brillouin phase choice
as $|\MW \pm \rangle _U ^{\wb}$.
Since
\beq 
| \pm \rangle _U ^{\wb} = |\MW \pm \rangle _U ^{\wb} 
\otimes |\MW \pm \rangle _U ^{\wb},
\eeq
we have
\beq
~^{\wb}_{~~U} \langle - | + \rangle _U ^{\wb} 
=
\left( ~^{\wb}_{~~U} \langle \MW \! - | \MW + \rangle _U ^{\wb} \right)^2 .
\label{eq:MWsquare}
\eeq
Thus the overlap formalism ensures that the fermion determinant
of a Weyl fermion is the square of that of a Majorana-Weyl fermion,
as it should be.

Our model can be defined through the following partition function.
\beq
Z = \int \prod_\mu [\dd U_\mu] \ee^{-S_b [U]}
~^{\wb}_{~~U} \langle \MW \! - | \MW + \rangle _U ^{\wb},
\eeq
where $[\dd U_\mu]$ denotes the Haar measure of the group integration
over U($N$). This gives a complete regularization of the IKKT model.

\subsection{Symmetries of the unitary IIB matrix model}

Let us turn to the symmetries of this model.
As is proved in Ref. \cite{NN},
the overlap has
several properties which are expected as a sensible lattice
regularization of a chiral determinant.
In the present case, we have
the invariance under 
10D discrete rotational transformation and the global gauge
transformation, among other things.
The invariance of the magnitude of the overlap is essentially because
the many-body Hamiltonians (\ref{eq:mbhamMW}) 
can be formally derived from 
(10+1)D Majorana fermion with masses of opposite signs.
The invariance of the phase comes from the fact that
the gauge configuration $U_\mu=1$, which is refered to in the 
Wigner-Brilloiun
phase choice, is invariant under these transformations.

Furthermore, we have the U(1)$^{10}$ symmetry:
\beq
U_\mu \rightarrow \ee^{i\alpha_\mu} U_\mu.
\eeq
This is because the boson action $S_b [U]$ and 
the Haar measure $[\dd U_\mu]$ is invariant,
and so is the fermion determinant given by the overlap,
since it depends only on $U_\mu^{\adj}$, which 
is invariant under the above transformation.

The fermion shift symmetry has its counterpart in the 
naive fermion action (\ref{eq:faction}),
in which 
the fermion component $\{\psi_0^{\adj},\bar{\psi}_0^{\adj}\}$ 
that corresponds to the 
generator $T^0={\bf 1}$ does not appear,
since 
\beq
(U_\mu^{\adj})_{00}=1;~~~~~
(U_\mu^{\adj})_{0a} = (U_\mu^{\adj})_{a0}=0
~~~~~\mbox{for}~ a \neq 0.
\label{eq:zeromode}
\eeq
$\psi_0^{\adj}$, which is the trace part of the fermionic
matrix, thus gives a zero mode.

We adopt this point of view in the overlap formalism,
since here the action is not defined, but only the determinant
after integrating over the fermionic matrices is given.
Due to eq.(\ref{eq:zeromode}), 
the component $\{\xi_0,\xi_0^\dagger\}$
that corresonds to the generator $T^0={\bf 1}$
decouples from the others in the many-body Hamiltonian as
\beqa
\mbham^{\mw} &= &
\frac{1}{2}
\left(
\begin{array}{cc}
\xi_0^{\dagger} & \xi_0^t
\end{array}
\right)
\left(
\begin{array}{cc}
\pm m & 0 \\
0 & \mp m
\end{array}
\right)
\left(
\begin{array}{c}
\xi_0  \\
\xi_0^{\dagger t}  
\end{array}
\right) + \cdots   \n
&=&  \pm m \prod_{s=1}^8  \xi_{0s} ^\dagger \xi_{0s} +  \cdots ,
\eeqa
where $s$ represents the spinor index.
Hence we can consider this part independently.
In the subspace on which $\xi_{0s}$ and $\xi_{0s}^\dagger$ act,
we define the kinematical vacuum $|0\rangle$ by $\xi_{0s} | 0 \rangle =
0$.
Then the ground states of the many-body Hamiltonians can be written as
\beqa
|\MW + \rangle _U ^{\wb} &=& | 0 \rangle \otimes \cdots \n
|\MW - \rangle _U ^{\wb} &=& \prod_{s=1}^8 \xi_{0s}^\dagger
| 0 \rangle \otimes \cdots . 
\eeqa
We, therefore, have
\beq 
~^{\wb}_{~~U} \langle \MW \! - | \MW + \rangle _U ^{\wb} = 0.
\eeq
Thus the trace part of the fermionic matrices gives a zero mode,
which means that the overlap has a symmetry that corresponds
to the fermion shift symmetry.
In order to obtain a non-zero expectation value out of this model,
we have to drop the zero-mode by hand, or equivalently, put them
in the observables.
In the rest of the paper, we assume that this is done implicitly,
as has been done in the IKKT model \cite{IKKT}.

To summarize, we have manifest symmetries corresponding to 
(iii),(iv) and (v).
On the other hand, the 10D continuous rotational group (i) 
is broken down to a discrete one and the zero-volume version 
of the supersymmetry (ii) is lost.
Thus, the situation is exactly the same as with lattice 
formulation of supersymmetric Yang-Mills theories.
As is advocated in Ref. \cite{superYM}, the overlap formalism can be used
to obtain all the supersymmetric Yang-Mills theories in 
the continuum limit without fine-tuning.
We expect here in the same spirit that 10D continuous rotational group 
(i) as well as the zero-volume version of the supersymmetry (ii) 
recovers in the double scaling limit without particular fine-tuning,
resulting in a theory with 10D ${\cal N}=2$ super Poincare invariance.

\subsection{The phase of the fermion determinant}

The fermion determinant in the IKKT model is complex in general,
and the phase depends on the bosonic matrices.
As is mentioned in the Appendix of Ref. \cite{IKKT}, however,
it is real in the following two cases.

Case (1) : $P_\mu = 0$ at least for one direction.

Case (2) : $F_{\mu\nu} = [P_\mu,P_\nu] = 0$ for $1 \le \mu < \nu \le
10$.

\noindent Correspondingly in the present model, 
the fermion determinant is complex
in general, and is real in the following two cases.

Case (1) : $U_\mu^{\adj} = 1$ at least for one direction.


Case (2) : $U_\mu^{\adj} U_\nu^{\adj} =U_\nu^{\adj} U_\mu^{\adj} $ 
for $1 \le \mu < \nu \le 10$.


\noindent We prove this in the following. 

Let us first consider the Case (1).
Without loss of generality, we can take $U_{10}^{\adj} = 1$.
Then we have
\beqa
\chiral &=& \sum_{i=1}^{9} \gamma_i D_i 
= \sum_{i=1}^{9} \gamma_i \frac{1}{2} (U_i^{\adj}-U_i^{\adj\dagger}), \n
\wilson &=& \sum_{i=1}^{9}
\left\{ 1 - \frac{1}{2}(U_i^{\adj} + U_i^{\adj\dagger}) \right\}.
\eeqa
The single-particle Hamiltonian
\beq
\ham = 
\left(
\begin{array}{cc}
\wilson \pm m & \chiral \Omega \\
(\chiral \Omega)^\dagger & -(\wilson \pm m)
\end{array}
\right)
\label{eq:hamMW}
\eeq
has the following symmetry.
\beq
\Sigma^\dagger \ham \Sigma = \ham ^*
\eeq
where 
\beq
\Sigma =
\left(
\begin{array}{cc}
\Omega & 0 \\
0 & \Omega^\dagger
\end{array}
\right).
\eeq
Therefore, the many-body Hamiltonian (\ref{eq:mbhamMW}) 
can be written as
\beq
\mbham^{\mw} = 
\frac{1}{2}
\left(
\begin{array}{cc}
\xi^{\dagger} & \xi^t
\end{array}
\right)
\ham 
\left(
\begin{array}{c}
\xi  \\
\xi^{\dagger t}  
\end{array}
\right)
=
\frac{1}{2}
\left(
\begin{array}{cc}
\eta^{\dagger} & \eta^t
\end{array}
\right)
\ham ^*
\left(
\begin{array}{c}
\eta  \\
\eta^{\dagger t}  
\end{array}
\right),
\eeq
where
$ \eta = \Omega \xi$.
The rest of the proof goes in exactly the same way as in 
Section III of Ref. \cite{superYM}, and we find that the 
fermion determinant defined through the overlap fromalism
is real in this case.
This result is naturally understood since
the fermion determinant considered in this case 
can be regarded as that of 
a massless pseudo-Majorana fermion in nine dimensions
\footnote{An application of the overlap formalism
to gauge theories in odd dimensions has been studied in Ref. 
\cite{parity,KikukawaNeuberger},
where the formalism can provide a parity invariant lattice regularization of 
massless Dirac fermion.
An extension to Majorana fermion is given in Ref. \cite{superYM}.
A similar construction can be made for pseudo-Majorana fermion in 9D.},
which is real.
One can further show that if $U_\mu ^{\adj} =1$ at least for four
directions, the overlap is not only real but also positive.
This is because the fermion determinant can now be considered as
that of massless Dirac fermion in six dimensions, which is real positive.

We next consider the Case (2).
Here we show that the overlap for a Weyl fermion is real positive,
which implies that the overlap for a Majorana-Weyl fermion
is real due to the relation (\ref{eq:MWsquare}).
Since $U_\mu^{\adj}$ commute with each other,
they can be diagonalized simultaneously as follows.
\beq
U_\mu^{\adj} = V^\dagger \Lambda_\mu V,
\eeq
where
\beq
(\Lambda_\mu)_{ab} = \ee^{i\theta_{\mu a}} \delta_{ab}.
\eeq
Due to this, the many-body Hamiltonians (\ref{eq:mbham}) 
can be decomposed as
\beqa
\mbham & = & 
\sum_a \mbham ^a \n
\mbham ^a & = &
\left(
\begin{array}{cc}
{\alpha_a '} ^\dagger & {\beta_a '} ^\dagger
\end{array}
\right)
\left(
\begin{array}{cc}
(b_a \pm m ){\bf 1} &  \chiral_a \\
\chiral_a^\dagger & - (b_a \pm m ){\bf 1}
\end{array}
\right)
\left(
\begin{array}{c}
\alpha_a '  \\
\beta_a '
\end{array}
\right),
\label{eq:decomp}
\eeqa
where we redefine the fermionic operators as
\beqa
\alpha_a ' &=& \sum_b V_{ab} \alpha_b \n
\beta_a ' &=& \sum_b V_{ab} \beta_b ,
\eeqa
and define
\beqa
\chiral_a & = & i \sum_{\mu} \gamma_\mu \sin \theta_{\mu a } \\
b_a &=& \sum_{\mu} (1 - \cos \theta_{\mu a })
= 2 \sum_{\mu} \sin ^2 \frac{\theta_{\mu a}}{2}.
\eeqa
The Hamiltonians $\mbham^a$ can be further decomposed
in the following way.
Note first that
$\sum_{i=1}^9 \gamma_i \sin \theta_{i a} $ is a hermitian matrix 
which has eigenvalues $\pm \sqrt{\sum_{i=1}^9 \sin ^2 \theta _{ia}}$, 
each of which has 8-fold degeneracy.
Therefore, $\chiral_a $ can be diagonalized as
\beq
\chiral_a = W_a ^\dagger
\left(
\begin{array}{cc}
z_a {\bf 1} &  0 \\
0 & z_a ^* {\bf 1}
\end{array}
\right)
W_a,
\eeq
where
\beq
z_a = - \sin \theta_{0a} + i \sqrt{\sum_{i=1}^9 \sin ^2 \theta_{ia}}.
\eeq
Redefining the fermionic operators as
\beqa
\left(
\begin{array}{c}
\phi_a  \\
\chi_a  
\end{array}
\right)
&=&
W_a
\alpha_a ' \\
\left(
\begin{array}{c}
\psi_a  \\
\omega_a  
\end{array}
\right)
&=&
W_a
\beta_a ',
\eeqa
we obtain
\beqa
\mbham^{a}&=& 
\left(
\begin{array}{cc}
\phi_a^\dagger & \psi_a^\dagger
\end{array}
\right)
\left(
\begin{array}{cc}
(b_a \pm m ) {\bf 1} & z_a {\bf 1} \\
z_a ^* {\bf 1}   & - (b_a \pm m ) {\bf 1} 
\end{array}
\right)
\left(
\begin{array}{c}
\phi_a  \\
\psi_a
\end{array}
\right)  \n
&~&+
\left(
\begin{array}{cc}
\chi_a ^\dagger & \omega_a ^\dagger
\end{array}
\right)
\left(
\begin{array}{cc}
(b_a \pm m ) {\bf 1}  & z_a ^* {\bf 1} \\
z_a {\bf 1}   & -(b_a \pm m) {\bf 1} 
\end{array}
\right)
\left(
\begin{array}{c}
\chi_a  \\
\omega_a
\end{array}
\right)  .
\eeqa

Thus the problem has reduced essentially to 
that of obtaining the overlap for the Hamiltonians
\beq
\widetilde{\mbham} = 
\left(
\begin{array}{cc}
c^\dagger & d^\dagger
\end{array}
\right)
\left(
\begin{array}{cc}
b \pm m  &  z \\
z^* & - (b \pm m )
\end{array}
\right)
\left(
\begin{array}{c}
c  \\
d
\end{array}
\right).
\eeq
The ground states $\widetilde{|\pm\rangle}_U ~\!\!\!\!\!\!^{\wb}$
of $\widetilde{\mbham}$ with the Wigner-Brillouin phase choice 
can be obtained as
\beqa
\widetilde{|+\rangle}_U ~\!\!\!\!\!\!^{\wb} &=&
\frac{1}{\sqrt{2 \epsilon^+ (\epsilon^+ + \mu^+ )}}
\left(  - z c ^\dagger + (\epsilon^+  + \mu^+) d ^\dagger  
\right) |0\rangle \\
\widetilde{|-\rangle}_U ~\!\!\!\!\!\!^{\wb} &=&
\frac{1}{\sqrt{2 \epsilon^- (\epsilon^- - \mu^- )}}
\left(
 (\epsilon^- - \mu^-) c ^\dagger
 - z^* d ^\dagger   \right) |0\rangle ,
\eeqa
where $\mu^\pm = b \pm m$ and 
$\epsilon^\pm = \sqrt{ |z|^2  + (\mu^\pm)^2  }$. 
$|0\rangle$ is the kinematical vacuum defined through
$ c |0\rangle = d |0\rangle = 0 $.
Note that $\widetilde{|+\rangle}_1 ~\!\!\!\!\!^{\wb}
 = d ^\dagger |0\rangle$ and
$\widetilde{|-\rangle}_1 ~\!\!\!\!\!^{\wb} = c ^\dagger |0\rangle$, 
and 
$~^{\wb} _{~~1}
\widetilde{\langle \pm } | \widetilde{\pm \rangle}
_U ~\!\!\!\!\!\!^{\wb}$ 
is real positive, which ensures the 
Wigner-Brillouin phase choice.
The overlap can be obtained as
\beq
~^{\wb} _{~~U}
\widetilde{\langle - } | \widetilde{+ \rangle}
_U ~\!\!\!\!\!\!^{\wb}
= \frac{-z}
{2 \sqrt{\epsilon^+ \epsilon^-} }
\left( \sqrt{\frac{\epsilon^- - \mu^-}{\epsilon^+ + \mu^+}}
+ \sqrt{\frac{\epsilon^+ + \mu^+}{\epsilon^- - \mu^-}}  \right).
\eeq

Now returning to our problem,
we find the overlap of the ground states of the many-body
Hamiltonians (\ref{eq:decomp}) as
\beq
~^{\wb} _{~~U}
\langle -  | + \rangle _U ^{\wb}
= \prod_a 
\left[
\frac{|z_a|^2}
{4 \epsilon^+ _a \epsilon^- _{a} }
\left( \sqrt{\frac{\epsilon^- _{a} - \mu^- _{a}}{\epsilon^+ _a + \mu^+ _a}}
+ \sqrt{\frac{\epsilon^+ _a + \mu^+ _a}{\epsilon^- _{a} - \mu^- _{a}}} 
 \right)^2 \right]^8,
\eeq
where $\mu_a^\pm= b_a \pm m$ and 
$\epsilon_a^\pm = \sqrt{ |z_a|^2 + (\mu_a^\pm)^2}$.
Thus the fermion determinant for a Weyl fermion
defined through the overlap formalism 
is real positive.
This completes the proof that the overlap for a Majorana-Weyl
fermion is real in the Case (2).

We can further show that the overlap for a Majorana-Weyl fermion is 
real positive in the Case (2) through the following argument.
We first note that since $\mbham ^{\mw}$ depend on $U_\mu^{\adj}$ 
continuously, we can choose the phases of the 
ground states $|\MW \pm \rangle _U $ so that the states depend
on $U_\mu^{\adj}$ continuously including their phases.
We define the ground states that obey the Wigner-Brillouin phase
choice as 
\beq
|\MW \pm \rangle _U ^{\wb} = \ee^{i \theta_{\pm} (U)}
|\MW \pm \rangle _U .
\eeq
Since
\beq
~^{\wb}_{~~1} \langle \MW \! \pm | \MW \pm \rangle _U ^{\wb} 
=
\ee ^{i( \theta _{\pm} (U) - \theta_{\pm}(1)) }
\!\!~_{~~1} \langle \MW \! \pm | \MW \pm \rangle _U ,
\eeq
the requirement that the left-hand side should be real positive
determines $\ee^{i \theta_{\pm}(U)}$ except when
$\!\!~_{~~1} \langle \MW \! \pm | \MW \pm \rangle _U = 0 $.
For configurations such that
$\!\!~_{~~1} \langle \MW \! \pm | \MW \pm \rangle _U \neq 0 $,
the phase factors $\ee^{i \theta_{\pm}(U)}$ 
and therefore the overlap
\beq
~^{\wb}_{~~U} \langle \MW \! - | \MW + \rangle _U ^{\wb} 
=
\ee ^{i ( \theta _{+} (U) - \theta_{-}(U))  }
\!\!~_{~~U} \langle \MW \! - | \MW + \rangle _U 
\eeq
are continuous functions of $U_\mu^{\adj}$.

The phase of 
$~^{\wb}_{~~U} \langle \MW \! - | \MW + \rangle _U ^{\wb} $
is therefore a continuous function of $U_\mu^{\adj}$ except when
\beqa
~^{\wb}_{~~1} \langle \MW \! \pm | \MW \pm \rangle _U ^{\wb} 
&=& 0,  \n
~^{\wb}_{~~U} \langle \MW \! - | \MW + \rangle _U ^{\wb} 
&=& 0.
\eeqa
Let us refer to such $U_\mu^{\adj}$ as {\it singular} configurations.
In the present case, $U_\mu^{\adj}$ is labeled by $\{ \theta_{\mu a} \}$
and the singular configurations correspond to the $\{ \theta_{\mu a} \}$
for which there exists an $a$ such that $\theta_{\mu a}=0$ or $\pi$ for
all $\mu$.
Note that any two configurations $\{ \theta_{\mu a} \}$ and
$\{ \theta_{\mu a} '\}$ which are not themselves singular
can be connected continuously without passing through singular 
configurations.
This means that the overlap can be chosen to be real positive for all
$\{ \theta_{\mu a} \}$
\footnote{Correspondingly in the IKKT model, a 
similar argument using continuity can be used to show
that the fermion determinant is real positive in the Case (2).}.

Therefore we can write the overlap for a Majorana-Weyl fermion as
\beq
~^{\wb}_{~~U} \langle \MW \! - | \MW + \rangle _U ^{\wb} 
= \prod_a 
\left[
\frac{|z_a|^2}
{4 \epsilon^+ _a \epsilon^- _{a} }
\left( \sqrt{\frac{\epsilon^- _{a} - \mu^- _{a}}{\epsilon^+ _a + \mu^+ _a}}
+ \sqrt{\frac{\epsilon^+ _a + \mu^+ _a}{\epsilon^- _{a} - \mu^- _{a}}} 
 \right)^2 \right]^4.
\eeq
Note also that the result depends only on
$b_a= 2 \sum_\mu  \sin ^2 \frac{\theta_{\mu a}}{2} $
and $|z_a|^2 = \sum_\mu \sin ^2 \theta _{\mu a }$,
which ensures the 10D discrete rotational invariance.
The explicit formula we have obtained here will be used in the 
next section to calculate the one-loop effective action around
some typical BPS-saturated configurations.

\section{One-loop effective action for BPS-saturated backgrounds} 
\setcounter{equation}{0}

In this section we calculate one-loop effective action around
some typical BPS-saturated backgrounds
such as $N$ D-instantons, one D-string and two parallel D-strings.
The classical equation of motion for the bosonic part can be
obtained by substituting 
$ U_\mu = \barU_\mu {\rm e}^{ia_\mu}$ in the action (\ref{eqn:S_b}) 
and requiring that the first order terms in $a_\mu$ cancel.
The result reads
\begin{equation}
\sum_{\nu}
(\barU_\nu W_{\mu\nu} \barU_\nu ^\dagger - W_{\mu\nu}) - \mbox{h.c.}=0
~~~~~\mbox{for all}~\mu,
\label{eqn:3000}
\end{equation}
where we define
\beq
W_{\mu\nu } = \barU_\mu ^\dagger \barU_\nu ^\dagger \barU_\mu \barU_\nu .
\eeq
By writing $\barU_\mu = \ee ^{i A_\mu}$ and expanding in 
the eq. (\ref{eqn:3000}) in terms of $A_\mu$,
we find to the leading order that
$$\sum_{\nu}[A_\nu,[A_\nu,A_\mu]]=0~~~~~\mbox{for all}~\mu,$$
which agrees with the equation of motion of the IKKT model
\cite{IKKT}.
One-loop effective action is calculated using the background field
method.
The contribution from the fermionic part is given by the overlap for
the background configuration $\bar{U}_\mu$.

We calculate the contribution from the bosonic part in the following.
The calculation can be done by a slight generalization of the one
given in Ref. \cite{BHN}.
We expand the bosonic unitary matrix around the background as
$$U_\mu =\barU_\mu \ee^{ia_\mu},$$
where $a_\mu$ is a hermitian matrix which represents the 
quantum fluctuation.
Putting this into the bosonic part of the action (\ref{eqn:S_b})
and expanding it up to the second order of $a_\mu$, we obtain
\beqa
S_b ^{(2)} &=&
- N \beta \sum_{\mu\nu}
\left[
\tr \left\{ 
(\barU_\nu ^\dagger a_\mu \barU_\nu - a_\mu )
(\barU_\mu ^\dagger a_\nu \barU_\mu - a_\nu ) W_{\mu\nu} \right\}
- \tr \left\{ [a_\mu, a_\nu] W_{\mu\nu} \right\}
\right. \n
&~&
~~~~~~~~~~~~
 - \frac{1}{2}
\tr \left\{ (\barU_\nu ^\dagger a_\mu \barU_\nu - a_\mu )
(\barU_\nu ^\dagger a_\mu \barU_\nu - a_\mu ) 
(W_{\mu\nu} + W_{\nu\mu}) \right\} \n
&~& 
~~~~~~~~~~~~
 \left. - \frac{1}{2}
\tr \left\{
 a_\mu [ \barU_\nu ^\dagger a_\mu \barU_\nu , 
(W_{\mu\nu} - W_{\nu\mu}) ] \right\} \right] .
\eeqa 
We use the following gauge fixing term
\beq
S_{g.f.} = N \beta
\tr \left\{  \sum_\mu
[\barU _\mu ^\dagger , U_\mu]
( \sum_\nu [\barU _\nu ^\dagger , U_\nu] ) ^\dagger
\right\}.
\eeq
Expanding this in term of $a_\mu$ up to the second order,
we obtain
\beq
S_{g.f.}^{(2)} =
N \beta \sum_{\mu\nu}
\tr \left \{ (\barU_\mu ^\dagger a_\mu \barU_\mu - a_\mu )
(\barU_\nu ^\dagger a_\nu \barU_\nu - a_\nu ) \right\}.
\eeq
The corresponding Faddeev-Popov ghost term is given by
\beq
S_{F.P.} = 
N \beta
\sum_\mu \tr \left\{ [b,\barU_\mu ^\dagger] [U_\mu,c] \right \}.
\eeq
Up to the second order of the quantum fields, we have
\beq
S_{F.P.} ^{(2)}= 
- N \beta
\sum_\mu 
\left[ \tr \left\{ 
b (\barU_\mu c \barU_\mu ^\dagger -c )
\right\}
+ \tr \left\{ 
b (\barU_\mu ^\dagger  c \barU_\mu  -c )
\right\} \right]
\eeq

The background configurations $\barU_\mu$ we consider in the following
correspond to BPS-saturated configurations for which 
$W_{\mu\nu} = \ee^{i \alpha _{\mu\nu}}$, where $\alpha_{\mu\nu}$
is either 0 or a c-number of order $\frac{1}{N}$.
Therefore $S_b ^{(2)}$ can be written as
\beqa
S_b ^{(2)} &=&
- N \beta \sum_{\mu\nu}
\cos \alpha_{\mu\nu}
\left[
\tr \left\{ (\barU_\mu ^\dagger a_\mu \barU_\mu - a_\mu )
(\barU_\nu ^\dagger a_\nu \barU_\nu - a_\nu ) \right\}  \right. \n
&~& ~~~~~~~~~~ - 
\left. \tr \left\{ (\barU_\nu ^\dagger a_\mu \barU_\nu - a_\mu )
(\barU_\nu ^\dagger a_\mu \barU_\nu - a_\mu ) \right\}
\right].
\eeqa 
Neglecting $O(\frac{1}{N^2})$, we can drop the factor 
$\cos \alpha_{\mu\nu}$.
Putting together, we obtain the total action for the bosonic
part up to the second order of the quantum fields as
\beqa
S^{(2)}
&=& S_b^{(2)} + S_{g.f.}^{(2)} + S_{F.P.}^{(2)} \n
&=& 
N \beta \sum_{\mu\nu}
\tr \left\{ (\barU_\nu ^\dagger a_\mu \barU_\nu - a_\mu )
(\barU_\nu ^\dagger a_\mu \barU_\nu - a_\mu ) \right\} \n
&~& 
- N \beta
\sum_\mu 
\left[ \tr \left\{ 
b (\barU_\mu c \barU_\mu ^\dagger -c )
\right\}
+ \tr \left\{ 
b (\barU_\mu ^\dagger  c \barU_\mu  -c )
\right\} \right] .
\label{eq:action2}
\eeqa
Introducing the adjoint notation through
\beq
(a_\mu)_{pq} = \sum_a (T^a)_{pq} (a_\mu ^{\adj})_a ,
\eeq
and similarly for the ghost fields,
we rewrite the eq. (\ref{eq:action2}) as
\beqa
S^{(2)}
&=& N \beta
\sum_\mu 
a_\mu ^{\adj t}
\sum_\rho (2 - \barU _\rho ^{\adj}  - \barU _\rho ^{\adj \dagger}  )
a_\mu ^{\adj} \n
&~& + N \beta
b^{\adj t}  \sum_\rho 
(2 - \barU _\rho ^{\adj}  - \barU _\rho ^{\adj \dagger}  ) c^{\adj},
\eeqa
where the adjoint link variable 
$\barU_\mu ^{\adj}$ is defined by
$ (\barU_\mu^{\adj})_{ab} = \tr (T^a \barU_\mu T^b 
\barU_\mu^\dagger)$.

The effective action can be obtained as
\beqa
W_b &=& - \log \int \dd a_\mu \dd b ~\dd c ~
\ee ^{- S ^{(2)}} \n
&=& 4 \log \det (\sum_\mu 
(2 - \barU _\mu ^{\adj}  - \barU _\mu ^{\adj \dagger}  )).
\eeqa
Since the $\barU_\mu^{\adj}$ commute with one another in the present case,
they can be diagonalized simultaneously, and we denote the
diagonal elements by $\lambda_{\mu a} = 
\ee ^{i \theta_{\mu a}}$.
Then we can write our result for the bosonic part as
\beq
W_b = 4 \sum_a 
\log (\sum _\mu 2 \sin ^2 \frac{\theta_{\mu a}}{2}).
\eeq
The fermionic part 
$W_f = - \log
~^{\wb}_{~~\barU} \langle \MW \! - | \MW + \rangle _{\barU} ^{\wb} $
is given by the formula obtained in the previous 
section, since the present case corresponds to the Case (2).
Adding the bosonic part and the fermionic part, we finally obtain
the total one-loop effective action as
\beqa
W_{tot} &=& W_b + W_f \n
&=&
4 \sum_a
\left[
\log b_a
-
\log
\left\{
\frac{\kappa_a}
{\epsilon^+ _a \epsilon^- _{a} }
\left( \sqrt{\frac{\epsilon^- _{a} - \mu^- _{a}}{\epsilon^+ _a + \mu^+ _a}}
+ \sqrt{\frac{\epsilon^+ _a + \mu^+ _a}{\epsilon^- _{a} - \mu^- _{a}} }
 \right)^2  \right\} \right],
\label{eq:final}
\eeqa
where
$\mu_a^\pm= b_a \pm m$, 
$\epsilon_a^\pm = \sqrt{ \kappa_a + (\mu_a^\pm)^2}$,
$b_a = 2\sum_\mu \sin ^2 \frac{\theta_{\mu a}}{2}$
and
$\kappa = \sum_\mu \sin ^2 \theta_{\mu a} $.
For each $a$ such that $\theta_{\mu a } = 0$ 
for all $\mu$, a zero mode appears
both in the bosonic part and in the fermionic part.
We should drop them by hand when we define the effective action.
In the summation over $a$ in eq. (\ref{eq:final}),
therefore, all such $a$'s should be excluded.
As can be seen from the above derivation,
ten times the number of zero modes is equal to the number of
flat directions of the bosonic action up to gauge transformation.
Note also that the $a$ that corresponds to $T^0 = {\bf 1}$
always gives a zero mode as 
is explained in section 2.3.
The corresponding flat directions are $\barU_\mu \rightarrow \barU _\mu
\ee ^{i\alpha_\mu}$.

In the following subsections, we consider specific BPS-saturated
backgrounds that corresponds to 
$N$ D-instantons, one D-string, and two parallel D-strings.
Since we have already obtained the general formula,
all we have to do is to diagonalize $\barU _\mu ^{\adj}$ explicitly
for each given background configuration $\barU_\mu$.

\subsection{Effective action for classcal vacua ($N$ D-instantons)}

In this subsection we consider classical vacua as the background
configuration.
The classical vacua, or the global minima of the bosonic action
(\ref{eqn:S_b}) is given by configurations $\barU_\mu$ which satisfy
$ \barU _\mu \barU _\nu = \barU _\nu \barU _\mu$.
Up to gauge transformation, we can take the following 
particular configurations.
\beq
\barU_\mu = \diag(\ee^{i\theta_{\mu 1}},\ee^{i\theta_{\mu 2}},\cdots,
\ee^{i\theta_{\mu N}}),
\eeq
where $\theta_{\mu I}$ ($I = 1,2,\cdots , N $) are arbitrary
parameters.
These configurations represent $N$ D-instantons, and the 
$\theta_{\mu I}$ are interpreted as 10D coordinates of the
$I$-th D-instanton.

In order to calculate $(\barU _\mu ^{\adj})_{ab}$,
we introduce the following explicit form for the generators $T^a$.
\beqa
(X^{(I,J),1})_{pq} &=& \frac{1}{\sqrt{2}}
(\delta_{pI} \delta_{qJ} + \delta_{pJ} \delta_{qI} ) \\
(X^{(I,J),2})_{pq} &=& \frac{1}{\sqrt{2}}
( - i \delta_{pI} \delta_{qJ} + i \delta_{pJ} \delta_{qI} ) \\
(Y^k)_{pq} &=& \delta_{pk} \delta_{qk}
\eeqa
where the indices $I$, $J$ and $k$ run over $1\le I < J \le N$
and $1 \le k \le N$, respectively.

$\barU_\mu^{\adj}$ is already block diagonal and has 
non-vanishing elements
only for the following choice of $a$ and $b$.

\noindent 
\underline{(i) $T^a=X^{(I,J),\alpha}$ and $T^b=X^{(I,J),\beta}$}

$(\barU_\mu^{\adj})_{ab} = (R_\mu^{(I,J)})_{\alpha\beta}$,
where the matrix $R_\mu^{(I,J)}$ is given as
\beq
R_\mu^{(I,J)} = \
\left(
\begin{array}{cc}
\cos (\theta_{\mu I} - \theta_{\mu J})   
& - \sin (\theta_{\mu I} - \theta_{\mu J} )   \\
\sin (\theta_{\mu I} - \theta_{\mu J} )   & 
\cos (\theta_{\mu I} - \theta_{\mu J})   
\end{array}
\right).                                                          \label{eqn:7}
\eeq
Therefore, the $(\barU_\mu^{\adj})_{ab}$ can be diagonalized simultaneously
and the diagonal elements $\lambda_{\mu}^{(I,J),\alpha}$ can be given as
\beq
\lambda_{\mu}^{(I,J),1}
=\ee^{i(\theta_{\mu I} - \theta_{\mu J})},~~~~~
\lambda_{\mu}^{(I,J),2}
=\ee^{-i(\theta_{\mu I} - \theta_{\mu J})}.
\label{eq:eigenDins}
\eeq

\noindent \underline{(ii) $T^a=Y^k$ and $T^b = Y^k$}

$(\barU_\mu)_{ab} =1$.
These give zero modes, which we subtract by hand.
Note that there are $N$ zero modes corresponding to 
$10N$ flat directions $\theta_{\mu I}$ for moving each of the 
$N$ D-instantons in arbitrary direction.

Substituting the $\lambda_\mu^{(I,J),\alpha}$ into the formula
(\ref{eq:final}), we obtain
\beqa
W^{(D-inst)} &=& 
\sum_{I < J} w _{tot} (I,J) \n
w _{tot} (I,J) &=& w_b (I,J) + w _f (I,J) \n
w_b (I,J) &=& 8 \log b_{IJ} \n
w _f (I,J) 
&=& -8 \log
\Biggr\{
\frac{\kappa_{IJ}}
{\epsilon^+ _{IJ} \epsilon^- _{IJ} }
\left( \sqrt{\frac{\epsilon^- _{IJ} - \mu^- _{IJ}}
{\epsilon^+ _{IJ} + \mu^+ _{IJ}}}
+ \sqrt{\frac{\epsilon^+ _{IJ} + \mu^+ _{IJ}}
{\epsilon^- _{IJ} - \mu^- _{IJ}}} 
 \right)^2 \Biggr\},
\eeqa
where we define $\mu^\pm_{IJ}= b_{IJ} \pm m$,
$\epsilon^\pm_{IJ} = \sqrt{
\kappa_{IJ} + (\mu^\pm_{IJ})^2 } $,
$b_{IJ}= 2 \sum_\mu 
\sin ^2 \left(\frac{\theta_{\mu I}- \theta_{\mu J}}{2} \right) $
and $ \kappa_{IJ} = \sum_\mu \sin ^2 (\theta _{\mu I}
-\theta _{\mu J } )$.
The result for the bosonic part agrees with the one obtained in
Ref. \cite{BHN}.

\begin{center}
\begin{minipage}[t]{165mm}
\begin{center}
\leavevmode\psfig{file=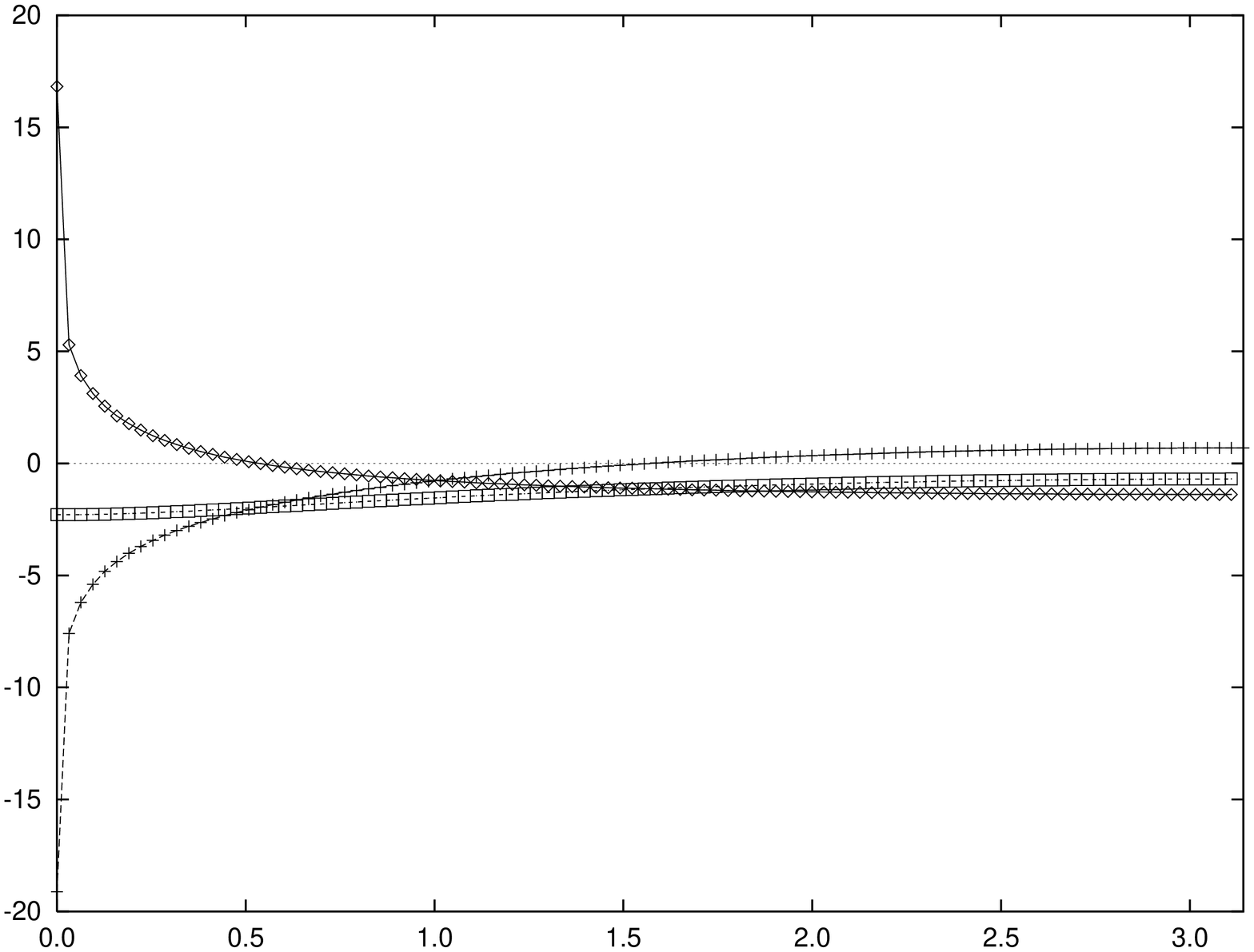,width=165mm,angle=270}
{\footnotesize Figure : The two-body potential between two instantons
is ploted against $|\theta_{1I} - \theta_{1J} |$.
The parameter $m$ is taken to be $m=0.9$. 
Crosses, diamonds and squares represent
$w_b$, $w_f$ and $w_{tot}$ respectively.}
\end{center}                                                    
 \label{fig:1}
\setlength{\baselineskip}{5mm}
\end{minipage}
\end{center}

The effective action is written as a sum of the two-body potential
$w_{tot}(I,J)$ between all the pairs of D-instantons.
In the Figure, we plot the two-body potential $w_{tot}(I,J)$
for the configuration with $\theta_{i I} - \theta_{i J} = 0 $
($i=2,3,\cdots,10$) as a function of $|\theta_{1 I} - \theta_{1 J}|$.
The parameter $m$ is taken to be 0.9, but the result is qualitatively
the same for any $m$ within $0<m<1$.
Note that the bosonic part gives an attractive potential,
while the fermionic part gives a repulsive one,
and the logarithmic singularity at
$|\theta_{1 I} - \theta_{1 J}| \approx 0$ in $w_b (I,J)$ and
$w_f (I,J)$ cancel each other.
Note also that another logarithmic singularity that
we would have in $w_f (I,J)$ at $|\theta_{1J} - \theta_{1J}|\approx  \pi$
if the naive ferimon action (\ref{eq:faction}) were used is 
absent thanks to the existence of $b_{IJ}$ in $\mu_{IJ}^\pm$.
This means that the doublers are successfully eliminated by the use
of the overlap formalism.
The total potential $w_{tot}(I,J)$ is monotonously increasing for 
$0 \le |\theta_{1 I} - \theta_{1 J}| \le \pi$ and the behavior
at small dimstance is given by
\beq
w_{tot} (I,J) \sim  \mbox{const.} + 8 
\left(\frac{1}{m^2} + \frac{1}{4}\right) 
| \theta_{1 I} - \theta_{1 J} |^2. 
\eeq
In the large $N$ limit, the D-instantons are attracted to each other
and the U(1)$^{10}$ symmetry is spontaneously broken.
Note, however, that the net attractive potential is much weaker
than the one for the purely bosonic case, which has logarithmic
singularity at the origin.

When one interprets the $\theta_{\mu I}$ as 10D coordinates of
the $I$-th D-instanton, the physical coordinates should be given by
\beq
X_{\mu I} = \frac{\theta_{\mu I}}{a},
\eeq 
where $a$ is a cutoff parameter which should be taken to zero keeping
$N a^2$ fixed when one takes the large $N$ limit according to 
Ref. \cite{FKKT}.
The two-body potential can be written in terms of the physical 
coordinates $X_{\mu I}$ as
\beq
w_{tot}(I,J) \sim \mbox{const.}+ 8 \left( \frac{1}{m^2}+ 
\frac{1}{4} \right)
a^2 \sum_\mu |X_{\mu I} - X_{\mu J}|^2 
+ O(a^4),
\eeq
and one might be 
tempted to consider that the potential is independent of 
$ \sum_\mu |X_{\mu I} - X_{\mu J}|^2 $ in the $a\rightarrow 0$ limit.
We should note, however, that the number of D-instantons $N$ 
goes to infinity as $a$ goes to zero.
If we introduce the density function $\rho(X)$ as
\beq
\rho (X) = \frac{1}{N} \sum_{I=1}^{N}
\prod_{\mu}^{10}
\delta(X_\mu - X_{\mu I}),
\eeq
the effective action can be written as
\beqa
W^{(D-inst)}
&=& N^2 \int \dd X_\mu ~ \dd Y_\mu  ~
\rho(X) 8 \left( \frac{1}{m^2}+ 
\frac{1}{4} \right)
a^2 \sum_\mu |X_{\mu} - Y_{\mu}|^2  \rho(Y) \n
&=& \mbox{const.}~N \int \dd X_\mu ~ \dd Y_\mu
~ \rho (X) \sum_\mu 
|X_{\mu} - Y_{\mu}|^2  \rho(Y) ,
\eeqa
where we have used the fact that we fix $N a^2$, when we take
the double scaling limit \cite{FKKT}.
Therefore, in the large $N$ limit, the above attractive potential 
dominates and the density function $\rho(X)$ approaches $\prod_\mu
\delta(X_\mu)$.
This means that although the two-body force between two D-instantons
placed at a finite physical distance
becomes weaker and weaker in the large $N$ limit,
the increase in the number of D-instantons results in
a collapse of the distribution function.
In this sense, our model is not equivalent to the IKKT model 
at least in the weak coupling limit.
We discuss further on this point in Section 4.

\subsection{Effective action for a D-string}

We next consider the background configuration which represents
one D-string.
\beq
\barU_1 = \Gamma_1 ;~~~~~ \barU_2 = \Gamma_2 ; ~~~~~\barU_i={\bf 1} 
~~~\mbox{for}~~~i=3,4,\cdots,10,                                 \label{eqn:10}
\eeq
where $\Gamma_1$ and $\Gamma_2$ are unitary matrices which satisfy 
the following algebra.
\beq
\Gamma_1 \Gamma_2 = \ee^{\frac{2\pi i }{N}} \Gamma_2 \Gamma_1.   \label{eqn:11}
\eeq

It is known that the solution to this algebra (\ref{eqn:11}) 
is unique up to the gauge transformation 
$\Gamma_j ' = g \Gamma_j g^\dagger $
and the U(1) transformation $\Gamma_j ' = \ee^{i \theta_j} \Gamma_j$,
where $j=1,2$.
Since our model is invariant under these transformations,
we take the following specific matrices without loss of generality.
\beqa
(\Gamma_1)_{pq} &=& \ee^{\frac{2 \pi i}{N} p} \delta_{pq} \n 
(\Gamma_2)_{pq} &=& \delta_{\overline{p+1},q} 
\eeqa 
We define the symbol $\overline{p}$ by
\beq
\overline{p} =
\left\{
\begin{array}{ll}
p  & \mbox{for}~~~l\le p \le N \\
p-N & \mbox{for}~~~N+1 \le p \le 2N.
\end{array}
\right.
\eeq
$\barU_\mu^{\adj}$ can be given by
\beqa
(\barU_1^{\adj})_{ab} 
&=& (T^a)_{pq} (T^b)_{qp} \ee^{\frac{2 \pi i}{N} (q-p)} \n
(\barU_2^{\adj})_{ab}
 &=& (T^a)_{pq} (T^b)_{\overline{q+1},\overline{p+1}} \n
(\barU_i^{\adj})_{ab} &=& \delta_{ab} ~~~~~\mbox{for}~~~i=3,4,\cdots,10.
\eeqa
In view of this, we introduce the following notation for the 
$X$-type generators.
\beqa
X^{Kn1} &=& X^{(n,\overline{K+n}),1},~~~
\mbox{{\it i.e.}},~~~(X^{Kn1})_{pq} =
\frac{1}{\sqrt{2}} (\delta_{pn} \delta_{q,\overline{K+n}} 
+\delta_{p,\overline{K+n}} \delta_{qn})
 \n
X^{Kn2} &=& X^{(n,\overline{K+n}),2}, ~~~
\mbox{{\it i.e.}},~~~(X^{Kn2})_{pq} =
\frac{1}{\sqrt{2}} ( -i \delta_{pn} \delta_{q,\overline{K+n}} 
+ i \delta_{p,\overline{K+n}} \delta_{qn})
\eeqa
where $n=1,2,\cdots,N$ and $K=1,2,\cdots, \frac{N-1}{2}$.
We assumed $N$ to be an odd number for simplicity.
One finds that $\barU_\mu^{\adj}$ is already block-diagonal
and has non-vanishing elements only for the following choice
of $a$ and $b$.

\noindent 
\underline{(i) $T^a = X^{Kn\alpha}$ and $T^b = X^{Km\beta}$}
\beqa
(\barU_1^{\adj})_{ab} &=& 
(R_K)_{\alpha\beta} \delta_{nm} \n
(\barU_2^{\adj})_{ab} &=& 
\delta_{\alpha\beta} \delta_{\overline{n+1},m} \n
(\barU_i^{\adj})_{ab} &=& \delta_{\alpha\beta} \delta_{nm}
~~~~~\mbox{for}~~~i=3,4,\cdots,10,
\eeqa
where the matrix $R_K$ is defined by
\beq
R_K = 
\left(
\begin{array}{cc}
\cos  \frac{2 \pi K }{N} & - \sin \frac{2 \pi K }{N} \\
 \sin \frac{2 \pi K }{N} & \cos  \frac{2 \pi K }{N} 
\end{array}
\right).
\label{eq:rotation}
\eeq
The $(\barU_\mu^{\adj})_{ab}$ can be diagonalized simultaneously
and the diagonal elements $\lambda_\mu ^{Kn\alpha}$ can be given as
\beqa
\lambda_1 ^{Kn\alpha} &=& 
\left\{
\begin{array}{ll}
\ee^{i\frac{2 \pi K}{N}} & \mbox{for}~~~\alpha=1 \\
\ee^{- i\frac{2 \pi K}{N}} & \mbox{for}~~~\alpha=2 
\end{array}
\right. \n
\lambda_2 ^{Kn\alpha} &=& \ee^{- i\frac{2 \pi n}{N}} \n
\lambda_i ^{Kn\alpha} &=& 1   ~~~~~\mbox{for}~~~i=3,4,\cdots, 10.\label{eqn:18}
\eeqa

\noindent \underline{(ii) $T^a = Y^k$ and $T^b = Y^l$}
\beqa
(\barU_1^{\adj})_{ab} &=& \delta_{kl} \n
(\barU_2^{\adj})_{ab} &=& \delta_{k,\overline{l+1}} \n
(\barU_i^{\adj})_{ab} &=& \delta_{kl} ~~~~~\mbox{for}~~~i=3,4,\cdots,10,
\eeqa
The $(\barU_\mu^{\adj})_{ab}$ can be diagonalized simultaneously
and the diagonal elements $\lambda_\mu ^{k}$ can be given as
\beqa
\lambda_1 ^{k} &=& 1 \n
\lambda_2 ^{k} &=& \ee^{i \frac{2 \pi k }{N}} \n
\lambda_i ^{k} &=& 1   ~~~~~\mbox{for}~~~i=3,4,\cdots, 10,
\label{eq:eigenDstr}
\eeqa
where $k=1,2,\cdots,N$.
$k=N$ gives the zero mode which we always have, 
and we drop it by hand.
The zero mode corresponds to the flat directions $\barU_\mu
\rightarrow \barU _\mu \ee^{i\alpha_\mu}$
for moving the D-string in arbitrary direction.

Putting all these eigenvalues into the formula (\ref{eq:final}),
we obtain
\beqa
W^{(1 D-str)}
&=&
8\sum_{K=1}^{\frac{N-1}{2}} \sum_{n=1}^{N}
\left[
\log b_{Kn} -
\log
\Biggr\{
\frac{\kappa_{Kn}}
{ \epsilon^+ _{Kn} \epsilon^- _{Kn} }
\left( \sqrt{\frac{\epsilon^- _{Kn} - \mu ^- _{Kn}}
{\epsilon^+ _{Kn} + \mu^+ _{Kn}}}
+ \sqrt{\frac{\epsilon^+ _{Kn} + \mu^+ _{Kn}}{\epsilon^- _{Kn} - \mu^- 
_{Kn}}} 
 \right)^2 \Biggr\} \right] \n
&~&
+4\sum_{k=1}^{N-1}
\left[ \log b_k - 
\log
\Biggr\{
\frac{\kappa_k}
{\epsilon^+ _k \epsilon^- _k }
\left( \sqrt{\frac{\epsilon^- _k - \mu ^- _k}
{\epsilon^+ _k + \mu^+ _k}}
+ \sqrt{\frac{\epsilon^+ _k + \mu^+ _k}{\epsilon^- _k - \mu^- 
_k}}  \right)^2 \Biggr\} \right],
\label{eq:dstring}
\eeqa
where we have defined
$\mu^\pm _{Kn}= b_{Kn} \pm m$,
$\epsilon^\pm _{Kn} = \sqrt{\kappa_{Kn} + (\mu^\pm _{Kn})^2 }$ and
\beqa
b_{Kn}  &=&  2 \left[ \sin ^2 \frac{\pi K}{N} + \sin ^2 \frac{\pi n}{N}
\right] \n
\kappa_{Kn}  &=& \sin ^2 \frac{2 \pi K}{N} + \sin ^2 \frac{2 \pi
n}{N}.
\eeqa
Similarly, we have defined
$\mu^\pm _k= b_k \pm m$,
$\epsilon^\pm _k = \sqrt{\kappa_k + (\mu^\pm _k)^2 }$ and
\beqa
b_k  &=&  2 \sin ^2 \frac{\pi k}{N} \n
\kappa_k &=& \sin ^2 \frac{2 \pi k}{N} .
\eeqa
The two terms in eq. (\ref{eq:dstring})
can be combined as follows by extending the
region of $K$ to $1\le K \le N$.
\beq
W^{(1 D-str)}
=
4\sum_{Kn} ~\!\!^{'}
\left[ \log b_{Kn}
- \log\Biggr\{
\frac{\kappa_{Kn}}
{ \epsilon^+ _{Kn} \epsilon^- _{Kn} }
\left( \sqrt{\frac{\epsilon^- _{Kn} - \mu ^- _{Kn}}
{\epsilon^+ _{Kn} + \mu^+ _{Kn}}}
+ \sqrt{\frac{\epsilon^+ _{Kn} + \mu^+ _{Kn}}{\epsilon^- _{Kn} - \mu^- 
_{Kn}}} 
 \right)^2 \Biggr\} \right], 
\eeq
where $\sum '_{Kn}$ denotes a summation over
$1\le K,n \le N $ except for $K=n=N$.
This result holds also when $N$ is an even number.

The effective action is equal to the one we would obtain
if D-instantons were broken into $N$ fractions,
and $N^2$ fractions made of $N$ D-instantons
were placed on the lattice sites on the surface swept by the
D-string. This picture is consistent with the fact that
the eigenvalues of the $\barU_1$ and $\barU_2$ are given by
$\ee^{i \frac{2 \pi p}{N}}$ where $p=1,2,\cdots,N$.
Physical length of the lattice spacing is $\frac{2\pi}{Na}$, which
goes to zero proportionally to $a$ when one takes the $N \rightarrow
\infty$ limit and $a \rightarrow 0$ limt with $N a^2$ fixed
\cite{FKKT}.
The physical extent of the space time, on the other hand,
is $\frac{2 \pi}{a}$, which goes to infinity at the same time.

\subsection{Effective action for two parallel D-strings}

Finally we consider a configuration which represents
two parallel D-strings.
\beqa
\barU_1 &=&
\left(
\begin{array}{cc}
\Gamma_1  &  \\
  & \Gamma_1
\end{array}
\right)   \n
\barU_2 &=&
\left(
\begin{array}{cc}
\Gamma_2  &  \\
  & 
\Gamma_2
\end{array}
\right)   \n
\barU_3 &=&
\left(
\begin{array}{cc}
\ee^{-i \frac{\theta}{2}} {\bf 1}   &  \\
  & \ee^{i \frac{\theta}{2}} {\bf 1}   
\end{array}
\right)
\eeqa
and $\barU_i=1$ for $i=4,\cdots,10$.
The matrices are now considered to be $2N\times 2N$.
We take the generators $T^a$ as follows.
\beqa
X_{(1)}^{Kn\alpha} &=& 
\left(
\begin{array}{cc}
X^{Kn\alpha} &  \\
  & 0
\end{array}
\right)  ; ~~~~~
Y_{(1)}^k = 
\left(
\begin{array}{cc}
Y^k &  \\
  & 0
\end{array}
\right)  \\
X_{(2)}^{Kn\alpha} &= &
\left(
\begin{array}{cc}
0 &  \\
  & X^{Kn\alpha}   \\
\end{array}
\right)  ; ~~~~~
Y_{(2)}^k = 
\left(
\begin{array}{cc}
0 &  \\
  & Y^k 
\end{array}
\right)  ,
\eeqa
where $K=1,2,\cdots, \frac{N-1}{2}$;
$n=1,2,\cdots,N$; $\alpha = 1,2$ and $k=1,2,\cdots,N$.
\beqa
(Z^{Kn1})_{pq} &=& \frac{1}{\sqrt{2}}
(\delta_{p, N+\overline{K+n}} \delta_{q,n} 
+ \delta_{p,n} \delta_{q,N+\overline{K+n}} ) \\
(Z^{Kn2})_{pq} &=& \frac{1}{\sqrt{2}}
(- i \delta_{p, N+\overline{K+n}} \delta_{q,n} 
+ i \delta_{p,n} \delta_{q,N+\overline{K+n}}
) ,
\eeqa
where $K=1,2,\cdots,N$ and $n=1,2,\cdots,N$.
$\barU_\mu^{\adj}$ is already block diagonal and has
non-vanishing elements
only for the following choice of $a$ and $b$.

\noindent \underline{(i) $T^a = X_{(i)}^{Kn\alpha}$ 
and $T^b = X_{(i)}^{Km\beta}$ ;
$T^a = Y_{(i)}^{k}$ 
and $T^b = Y_{(i)}^{l}$}

These give self-energy of each D-string obtained in the previous 
subsection. There is a zero mode for each D-string,
which corresponds to the flat directions 
\beq
\barU_\mu \rightarrow \barU_\mu  
\left(
\begin{array}{cc}
\ee^{i \alpha_\mu} {\bf 1}   &  \\
  & \ee^{i \beta_\mu} {\bf 1}   
\end{array}
\right)
\eeq
for moving the two D-strings in arbitrary direction separately.
Although the result should be invariant for 
moving the two D-strings together in the same direction
due to the U(1)$^{10}$ symmetry of the model,
it is not necessarily so for moving them in different directions.
Among such moves we are concentrating on the one
given by the $\theta$ in $\barU_3$.
It is straightfoward to redo the calculation including the other nine 
parameters in the background configuration.

\noindent \underline{(ii) $T^a=Z^{Kn\alpha}$ 
and $T^b=Z^{Km\beta}$}

This gives the interaction energy between the two parallel D-strings.
\beqa
(\barU_1^{\adj})_{ab} &=& 
(R_K)_{\alpha\beta} \delta_{nm} \n
(\barU_2^{\adj})_{ab} &=& 
\delta_{\alpha\beta} \delta_{\overline{n+1},m} \n
(\barU_3^{\adj})_{ab} &=& R(\theta)_{\alpha\beta} \delta_{nm} \n
(\barU_i^{\adj})_{ab} &=& \delta_{\alpha\beta} \delta_{nm}
~~~~~\mbox{for}~~~i=4,\cdots,10,
\eeqa
where the matrix $R_K$ is given by eq. (\ref{eq:rotation})
and $R(\theta)$ is defined by
\beq
R(\theta) = 
\left(
\begin{array}{cc}
\cos  \theta & - \sin \theta \\
 \sin \theta & \cos  \theta
\end{array}
\right)
\eeq
The $(\barU_\mu^{\adj})_{ab}$ can be diagonalized simultaneously
and the diagonal elements $\lambda_\mu ^{Kn\alpha}$ can be given as
\beqa
\lambda_1 ^{Kn\alpha} &=& 
\left\{
\begin{array}{ll}
\ee^{i\frac{2 \pi K}{N}} & \mbox{for}~~~\alpha=1 \\
\ee^{- i\frac{2 \pi K}{N}} & \mbox{for}~~~\alpha=2 
\end{array}
\right. \n
\lambda_2 ^{Kn\alpha} &=& \ee^{i\frac{2 \pi n}{N}} \n
\lambda_3 ^{Kn\alpha} &=& 
\left\{
\begin{array}{ll}
\ee^{i  \theta} & \mbox{for}~~~\alpha=1 \\
\ee^{- i \theta} & \mbox{for}~~~\alpha=2 
\end{array}
\right. \n
\lambda_i ^{Kn\alpha} &=& 1 ~~~~~\mbox{for}~~~i=4,\cdots, 10.
\label{eqn:40}
\eeqa

Substituting the $\lambda_\mu^{Kn\alpha}$ into the formula
(\ref{eq:final}), we obtain
\begin{eqnarray*}
W^{(2 par. D-str)}&=&2W^{(1 D-str)}+W^{(int)},  \\
W^{(int)}&=&
8\sum_{K,n}\Biggr[
\log b_{Kn}(\theta) \n
&~& ~~~~~
 - \log\Biggr\{
\frac{\kappa_{Kn}(\theta)}
{\epsilon^+ _{Kn}(\theta)  \epsilon^- _{Kn}(\theta) }
\left( \sqrt{\frac{\epsilon^- _{Kn}(\theta) - \mu^- _{Kn}(\theta)}
{\epsilon^+ _{Kn}(\theta) + \mu^+ _{Kn}(\theta) }}
+ \sqrt{\frac{\epsilon^+ _{Kn}(\theta) + \mu^+ _{Kn}(\theta)}
{\epsilon^- _{Kn}(\theta) - \mu^- _{Kn} (\theta)}} 
 \right)^2 \Biggr\}   \Biggr],
\end{eqnarray*}
where $\mu^\pm_{Kn} (\theta)= b_{Kn} (\theta)\pm m$ and 
$\epsilon^\pm_{Kn} (\theta) = \sqrt{
\kappa_{Kn}(\theta) + (\mu^\pm_{Kn}(\theta))^2 }$.
$ b_{Kn} (\theta)$ and 
$\kappa_{Kn}(\theta) $
are given by
\beqa
b_{Kn}(\theta) &=& 
2 \left[ \sin ^2 \frac{\pi K}{N} + \sin ^2 \frac{\pi n}{N}
 + \sin ^2 \frac{\theta}{2}
\right] \n
\kappa_{Kn}(\theta) &=& 
\sin ^2 \frac{2 \pi K}{N} + \sin ^2 \frac{2 \pi n}{N}
 + \sin ^2 \theta .
\eeqa

When we make a physical interpretation of this result, we 
have to fix the physical distance between the D-strings given by
$X = \frac{\theta}{a}$
and 
divide the effective action by the interaction time
$T = \frac{2\pi }{a}$ and the length of the D-strings
$L = \frac{2\pi }{a}$ to get the interaction energy between the 
two D-strings per unit length.
\beq
\varepsilon = \frac{W^{(int)}}{TL} \sim a^4 N^2 |X|^2.
\eeq
Therefore, in the large $N$ limit with $N a^2$ fixed as before, 
we have $\varepsilon \sim \mbox{const.} |X|^2$,
which depends on the physical distance of the two parallel D-strings.
This again shows that our model is not equivalent to the
IKKT model in the weak coupling limit.

\section{Double scaling 
limit and the dynamical generation of the space time}
\setcounter{equation}{0}

In order to reproduce string theories from matrix models,
we have to take the double scaling limit.
In our model,
the U(1)$^{10}$ symmetry is spontaneously broken in the 
weak coupling limit and is 
expected to be restored in the
strong coupling regime, which means that there must be a phase
transition
somewhere in between.
It is natural to identify the critical point of 
this phase transition associated with
the spontaneous breakdown of the U(1)$^{10}$ symmetry as
the place where we take the double scaling limit.
Let us consider approaching the critical point from the strong
coupling regime.
Since it is natural to relate our model to the IKKT model through
$U_\mu = \ee^{i a A_\mu}$,
we obtain to the leading order of $a$ as 
\beq
S = - \frac{1}{2} N \beta a^4 \tr [A_\mu, A_\nu]^2 + O(a^6).
\eeq
We therefore have $\frac{1}{2 g^2}= N\beta a^4$.
In Ref. \cite{FKKT}, it has been shown through the study of 
Schwinger-Dyson equation that the double scaling limit is 
given by fixing
\beqa
g^2 N &=& {\alpha '} ^2 \\
N a^2 &=& \frac{1}{g_{str} \alpha '},
\eeqa
when one takes the large $N$ limit,
where $g$ is the coupling constant in the action (\ref{eq:IKKTaction}).
Translating this into our model, we have
\beqa
\frac{\beta}{N^2} &=& \frac{1}{2} g_{str}^2 \\
\beta a ^4 &=& \frac{1}{ 2 {\alpha '}^2},
\eeqa
which means that $\beta$ has to be sent to infinity as $\beta \sim N^2$.
In order to approach the critical point in the double scaling limit,
the pseudo-critical point $\beta_c (N)$ for finite $N$ 
must behave in the large $N$ limit as
$\beta_c(N) \sim N^2$.
This gives a nontrivial test of our scenario which could be checked
by numerical simulation in principle.
This rather unusual shift of the critical point is not very unlikely,
considering the fact that actually the U(1)$^{10}$ symmetry is much
more mildly broken in the weak coupling limit than in the purely
bosonic case where there exists an attractive logarithmic
potential among the D-objects and the spontaneous breakdown 
of the U(1)$^{D}$ symmetry 
occurs at finite $\beta$ in the large $N$ limit.
The nontrivial cancellation of the logarithmic potentials from
the bosonic part and the fermionic part is actually due to 
the supersymmetry
at $U_\mu \sim 1$, where our model reduces to the IKKT model.
In this sense, although supersymmetry is not manifest in our model,
it plays an important role in the above argument.

The results in the weak coupling limit for our model 
are different from the ones for the IKKT model,
which have been regarded as an evidence that the IKKT model
reproduces the massless spectrum of the type IIB superstring theory
in Ref. \cite{IKKT}.
We point out, however, that the calculation in the weak coupling limit
has nothing to do with the double scaling limit {\it a priori}.
In this sense, we regard the success of the IKKT model 
in the weak coupling limit as accidental.
More to the point, we consider that the fact that the U(1)$^{10}$
symmetry is not spontaneously broken in the IKKT model even in the 
weak coupling limit suggests that it is never broken spontaneously
throughout the whole region of the coupling constant,
which we suspect is the reason why
the properties that should be attributed to the double scaling limit
have been obtained even in the weak coupling limit.
In the present model, on the other hand,
if the pseudo-crititcal point $\beta_c (N)$ for finite $N$
goes to infinity in the large $N$ limit as is claimed above,
the $\beta\rightarrow\infty$ limit and the $N\rightarrow\infty$ limit
is not commutable and the result should depend on how one takes the 
two limits.
Therefore the results in the weak coupling limit do not necessarily 
reflect the properties in the double scaling limit.

The spontaneous breakdown of the U(1)$^{10}$
symmetry in the weak coupling limit is actually welcome
when we consider the dynamical generation of 
the four-dimensional space time in the double scaling limit.
Let us speculate on 
how we could hope for obtaining the four-dimensional 
space time through our model in the double scaling limit.
Since the phase transition is associated with spontaneous breakdown
of the U(1)$^{10}$ symmetry, which is interpreted as the translational 
invariance of the 10D space time, naturally we would have a space time
of dimension between 0 and 10 in the double scaling limit.
We give a hand-waving argument for obtaining the space-time
dimension four in the double scaling limit for sufficiently small
$g_{str}$.
Here, planar diagrams dominate and the equivalence of the reduced
model and the large $N$ gauge theory is expected due to the argument
of Eguchi-Kawai \cite{EK}.
Rougly speaking, if U(1)$^{10}$ symmetry is broken down to 
U(1)$^D$, the reduced model is equivalent to $D$-dimensional gauge
theory.
We pay attention to the fact 
that the critical dimension of gauge theories is four.
Namely, we can obtain a nontrivial 
continuum theory by approaching the Gaussian
fixed point only when the dimension of the space time is equal to 
or less than four.
This suggests that the $D$ of the remaining U(1)$^D$ symmetry
should be equal to or less than four
\footnote{In Ref. \cite{largeN}, an essential difference in the phase
diagram between the one-site model with U(1)$^4$ symmetry 
and the one with U(1)$^6$ symmetry has been revealed for purely
bosonic case through Monte Carlo simulation.}.
We do not have any plausible
reasoning to exclude the possibility of the space-time 
dimenion turning out to be less than four, 
but naively we may expect that the 
critical value ``four'' has a special meaning.

\section{Summary and Future Prospects}
\setcounter{equation}{0}

In this paper, we proposed a unitary model as a regularization 
of the IKKT model, which is considered to give a nonperturbative
definition of the type IIB superstring theory.
Our model preserves manifest U(1)$^{10}$ symmetry, which corresponds
to the ten-dimensional translational invariance.
On the other hand, the ${\cal N}=2$ supersymmetry as well as
the continuous 10D Lorentz invariance is expected to be restored only
in the double scaling limit.

One-loop calculation of the effective action around some typical
BPS-saturated states has been performed.
The corresponding calculation in the IKKT model is done by
formally taking $N = \infty$ and $a =0$ from the beginning,
while we have done our calculation for finite $N$ and non-vanishing
$a$.
The results for our model differ from the ones for the IKKT model.
Above all, the U(1)$^{10}$ symmetry is spontaneously broken in our
model.
We argued, however, that this is not a problem itself.
Rather, the phase transition associated with the U(1)$^{10}$ symmetry
breakdown provides a natural place to take the double scaling limit.
The crucial feature necessary for our model to work is 
that the pseudo-critical coupling $\beta_c(N)$ shifts as $N^2$ when one
takes $N$ to the infinity.
The spontaneous U(1)$^{10}$ symmetry breakdown is also welcome for
a natural explanation 
of the dynamical generation of the space time in the double scaling limit.
We gave a hand-waving argument 
that for sufficiently small $g_{str}$, 
the space-time dimension is likely to be four.

Monte Carlo simulation of our model and the IKKT model
is possible in principle.
We hope that it will show whether our considerations are correct or
not.
It would be interesting if we could make some approximation 
and extract some nonperturbative physics analytically,
just as in Ref. \cite{Wilson}
the qualitative understanding of the quark confinement has been given
by the strong coupling expansion.

\vspace{1cm}

The authors would like to thank M. Fukuma, T. Izubuchi, H. Kawai, 
M. Kawanami, Y. Matsuo, R. Narayanan and A. Tsuchiya 
for valuable discussions and comments.

\end{document}